\documentclass[prd,aps,twocolumn,showpacs,preprintnumbers,nofootinbib,superscriptaddress,notitlepage]{revtex4-1}
\usepackage{epsfig}
\usepackage{graphicx}
\usepackage{bm}
\usepackage{times}
\usepackage{braket}
\usepackage{color}
\usepackage{slashed}
\usepackage{hyperref}
\usepackage{ulem}

\graphicspath{{./figure/}}

\definecolor{Red}{rgb}{1.,0.,0.}

\definecolor{Blue}{rgb}{0.,0.,1.}

\definecolor{Green}{rgb}{0.,1.,0.}

\definecolor{Gray}{rgb}{0.5,0.5,0.5}

\definecolor{nicered}{rgb}{0.7,0.1,0.1}
\definecolor{nicegreen}{rgb}{0.1,0.5,0.1}

\newcommand{\beq}{\begin{eqnarray}}
\newcommand{\eeq}{\end{eqnarray}}
\newcommand{\non}{\nonumber\\ }

\newcommand{\zsl}{ z \hspace{-2.2truemm}/ }

\hypersetup{colorlinks,citecolor=nicegreen,linkcolor=nicered}

\begin{document}


\title{Determination of pion distribution amplitudes from the electromagnetic form factor with the data-driven dispersion relation}

\author{Jian Chai}  
\affiliation{School of Physics and Electronics, Hunan University, 410082 Changsha, China.}
\author{Shan Cheng}\email{scheng@hnu.edu.cn}
\affiliation{School of Physics and Electronics, Hunan University, 410082 Changsha, China.}
\affiliation{Hunan Provincial Key Laboratory of High-Energy Scale Physics and Applications, 410082, Changsha, China.}
\author{Jun Hua}\email{junhua@scnu.edu.cn}
\affiliation{Guangdong Provincial Key Laboratory of Nuclear Science, Institute of Quantum Matter,
South China Normal University, Guangzhou 510006, China.}
\affiliation{Guangdong-Hong Kong Joint Laboratory of Quantum Matter,
Southern Nuclear Science Computing Center, South China Normal University, Guangzhou 510006, China.}

\date{\today}

\begin{abstract}

We study the pion electromagnetic form factor in the modulus squared dispersion relation, 
and do the model independent extraction of the most important nonperturbative parameters in pion light-cone distribution amplitude. 
The motivation of this work is the recent measurement of timelike form factor in the resonant regions, 
which makes up the piece lacking solid QCD-based calculation. 
With the perturbative QCD calculation up to next-to-leading-order QCD corrections and twist four level of meson distribution amplitudes, 
we obtain the chiral mass of pion meson as $m_0^\pi(1 \, {\rm GeV}) = 1.31^{+0.27}_{-0.30} \, {\rm GeV}$.  
More accuracy measurement of the timelike form factor are desirable to help us to determine the lowest gegenbauer moments 
in pion distribution amplitudes with the data-driven dispersion relation method developed in this work.

\end{abstract}



\maketitle 

\section{Introduction}

The calculation of matrix element is the pivotal issue to be addressed in a physical process involved hadrons. 
As the simplest physical observable corresponding to hadron matrix element, 
pion electromagnetic  form factors play an indispensable role in the QCD study, such as the development of factorization theorem, 
and the investigation of hadron structure \cite{Lepage:1980fj,Efremov:1979qk}.  

Electromagnetic (EM) form factor describes the interaction strength of momentum redistribution in a hadron 
when it is hinted by an energetic photon whereas not breaks up.  
To our knowledge, it can be calculated by four different QCD-based approaches, 
saying the lattice QCD (LQCD), Dyson-Schwinger equation (DSE), Light-cone sum rules (LCSRs) and the perturbative QCD (pQCD) approaches. 
These approaches realize the full predictions of the form factor in the spacelike regions. 
Speaking in turns, the LQCD recently have improved their evaluation ability in the region $- 1 \, {\rm GeV}^2 \leq q^2 \leq 0$ \cite{Wang:2020nbf}, 
the DSE is applicable in the low momentum transfers $-5 \, {\rm GeV}^2 \leq q^2 \leq -1 \,{\rm GeV}^2$ \cite{,Chang:2013nia,Roberts:2021nhw}, 
the LCSRs calculation based on operator production expansion is valid in the low and intermediate momentum transfers 
$-10 \, {\rm GeV}^2 \leq q^2 \leq -1 \, {\rm GeV}^2$ \cite{,Braun:1999uj,Cheng:2020vwr}, 
and the pQCD applies to the process with large momentum transfer $\vert q^2 \vert, q^2 \gtrsim 10 \, {\rm GeV}^2$ \cite{Jain:1999xc,Cheng:2019ruz}.
From the experiment side, the spacelike form factor is usually measured via the electron-nucleon elastic scattering \cite{NA7:1986vav} 
and the electron produced pion meson process $^1H(e,e'\pi^+)n$ \cite{JeffersonLabFpi-2:2006ysh,JeffersonLab:2008jve}, 
however, the precise result are obtained only with the small momentum transfers $-2.50 \, {\rm GeV}^2 \leq q^2 \leq -0.25 \,{\rm GeV}^2$. 
Meanwhile, the timelike form factor have been measured at $B$ factories with high accuracy. 
For examples, the isospin-vector form factor is measured via the $\tau$ decays in the momentum transfer region 
$4m_\pi^2 \leq q^2 \leq 3.125 \,{\rm GeV}^2$ by the Belle collaboration \cite{Belle:2008xpe}, 
and via the $e^+e^-$ annihilation process in the region $4 m_\pi^2 \leq q^2 \lesssim 8.7 \, {\rm GeV}^2$
by the BABAR collaboration \cite{BaBar:2012bdw}. 
What's more, the BESIII collaboration have also reported the precise result on the timelike form factor in the low momentum transfer 
$0.6 \,{\rm GeV}^2 \leq q^2 \leq 0.9 \,{\rm GeV}^2$ based on the initial state radiation (ISR) method \cite{BESIII:2015equ}. 
With the possible intermediate resonants and their interactions, 
these precise measurements could not be explained by the direct perturbative QCD calculation. 

In this work, we focus on the pQCD study based on $k_T$ factorization theorem \cite{Li:1992nu}, 
in which the transversal momentum is picked up to regularize the end-point divergence and 
the resummation techniques are embodied to suppress the large logarithms generated by gluon radiations. 
In this way, the processes with large momentum transfers are dominant conducted by the hard scattering 
and hence calculable in perturbative theorem. 
There are three uncertainty sources in pQCD calculation, besides the frequently-studied QCD radiation corrections \cite{Li:2010nn,Cheng:2014gba,Hu:2012cp,Cheng:2015qra} and the power corrections arose from light-cone distribution amplitudes (LCDAs) and interaction operators \cite{Chen:2018tch,Shen:2019zvh}, 
the theoretical-self uncertainty needs equal attention to improve the prediction power. 
This uncertainty mainly comes from the input parameters of hadrons and the choice of the factorization and normalization scale. 
In the pQCD calculation of pion EM form factor, 
the apparent skewed distribution of subleading twist LCDAs results in the so called chiral enhancement effect, 
saying, the contribution from twist three LCDAs is larger than that from the leading twist LCDA in the non-ultraviolet region. 
The subleading twist terms are proportional to the chiral mass $m_0^\pi$ whose value engenders the largest uncertainty in the pQCD prediction. 
Meanwhile, the non-asymptotic terms in the leading twist contributions also engender sizable uncertainty 
especially in the regions with low and intermediate momentum transfers, 
it is originated from the choice of Gegenbauer moments and mainly from the second moment $a_2^\pi$. 
Concerning the factorization $\mu_f$ and renormalization scales $\mu_R$ in the pQCD calculation, 
the convention choice is both at the largest virtuality in the hard scattering $\mu_t = {\rm Max}(x_iQ^2, 1/b_i)$ 
on the basis of the typical scale determined by external variables $Q^2$ and $k_{iT}^2$, 
here $x_i, b_i$ denote the longitudinal momentum fraction carried by a parton and the transversal interval extended in a meson, respectively. 
In the pion EM radiative process, the typical scales are $x_i Q^2 \pm k_{iT}^2$ in the spacelike/timelike transitions, 
in this way, the conventional choice is still artificial since other choices 
between the extreme alternatives $\mu = Q^2 \pm k_{iT}^2$ and $\mu = k_{iT}^2$ are all reasonable.

The main target of this paper is to extract the most important nonperturbative parameters $m_0^\pi$ and $a_2^\pi$ in the study of pion EM form factor. 
To archive it, we employ the dispersion relation in which the timelike form factor in the resonant regions have been measured by BABAR collaboration, 
and the form factors in the timelike and spacelike regions with large invariant masses and momentum transfers, respectively, 
are well calculated by the pQCD approach. 
Our calculation takes into account all the current known next-to-leading-order (NLO) corrections, 
and the accuracy of power expansion is up to twist four both for two-particle and three-particle LCDAs. 
We consider the renormalization evolution of nonperturbative parameters at the variable scale $\mu_t$, 
rather than the conventional choice at the default scale $\mu_0 = 1 \,{\rm GeV}$. 
Besides the errors from the BABAR measurements, we also examine the influence from the scale choice in the pQCD calculation. 

The paper is arranged as follow. In the next section, the dispersion relations are introduced in the standard and modified formalisms, 
In section \ref{sec:pQCD-form}, the pQCD formalism is demonstrated. 
We then do the fit between the spacelike form factor obtained from the dispersion relation and the direct pQCD calculation in section \ref{sec:fit}.  
The summary is given in section \ref{sec:summary}.

\section{Dispersion relations}\label{sec:ff_DR}

With considering the analytical properties follow from causality and Cauchy's theorem, 
the real and imaginary parts of scattering amplitudes are related to each other by the dispersion relation. 
In light of this, the full pion EM form factor in the $q^2 < 4m_\pi^2$ regions 
can be written as an integral over its imaginary parts with the unsubtracted formula \cite{Donoghue:1996kw,Zwicky:2016lka}
\beq
{\cal F}_\pi(q^2) = \frac{1}{\pi} \int_{4m_\pi^2}^\infty ds \frac{{\rm Im}{\cal F}_\pi(s)}{s - q^2 - i \epsilon} \,.
\label{eq:Fpi_DR1}
\eeq
This relation could reproduce the QCD asymptotics of the pion form factor ${\cal F}_\pi(q^2) \sim 1/q^2$ at $q^2 \rightarrow \infty$ in the power limit \cite{Lepage:1980fj,Efremov:1979qk}, that is why the subtractions are not necessary in the dispersion relation of pion EM form factor.

The measurement of timelike form factor in the right hand side (RHS) of Eq. (\ref{eq:Fpi_DR1}) 
is carried out for the modulus square $\vert {\cal F}_\pi(s) \vert^2$, 
and then the information of imaginary part is usually obtained by a parameterization of ${\cal F}_\pi(s)$, 
such as the Gounaris-Sakurai (GS) representation \cite{Gounaris:1968mw} and the K\"uhn-Santamaria (KS) representation \cite{,Kuhn:1990ad}.  
In this sense, the standard dispersion relation in Eq. (\ref{eq:Fpi_DR1}) brings an inevitable model dependence 
which results in an additional uncertainty to the spacelike form factor on the left hand side (LHS). 
In order to get rid off the model dependence in describing the timelike form factor at low and intermediate momentum transfers, 
the dispersion integral is modified in the modulus squared formalism \cite{Cheng:2020vwr}, 
\beq
\mathcal{F}^{\rm DR}_\pi (q^2) &=& \exp \left[ \frac{q^2 \sqrt{s_0 - q^2}}{2 \pi} \int\limits_{4m_\pi^2}^\infty d s 
\frac{ \ln \vert \mathcal{F}_\pi (s) \vert^2}{s\,\sqrt{s - s_0}  \, (s -q^2)} \right] \,. 
\label{eq:Fpi_DR2}
\eeq
The modulus square of timelike form factor is written in terms of heavy theta functions 
to separate the data measured in the region $[4m_\pi^2, s_{\rm max} \simeq 8.7 \, {\rm GeV}^2]$ and the high energy tail, 
\beq
\vert \mathcal{F}_{\pi}(s) \vert^2 &=& \Theta(s_{\rm max} - s) \vert \mathcal{F}^{\rm data}_{\pi}(s) \vert^2 \nonumber\\
&+& \Theta(s - s_{\rm max}) \vert \mathcal{F}_{\pi}^{\rm tail} (s) \vert^2 \,.
\label{eq:Fpi_DR2_tl_1}
\eeq 

The modified formalism in Eq. (\ref{eq:Fpi_DR2}) improves the accuracy of dispersion relation 
by skipping the reconstruction of imaginary part of timelike form factor, 
while the special expressions of $\vert \mathcal{F}_{\pi}(s) \vert^2$ in Eq. (\ref{eq:Fpi_DR2_tl_1}) could bring additional model dependence. 
For example, in the LCSRs work \cite{Cheng:2020vwr}, the piece of data was parameterized in the GS model 
by taking in to account the contributions from $\rho, \rho^\prime, \rho^{\prime\prime}$ and $\omega$ \cite{BaBar:2012bdw}, 
and the high energy tail beyond the experiment availability was written in the duality resonant model (DRM) \cite{Dominguez:2001zu,Bruch:2004py}. 
In this work, we study the modified dispersion relation from the view of pQCD approach. 
An unique advantage here is that the high energy tail can be directly calculated by perturbative theorem. 
For the measurement \cite{BaBar:2012bdw}, the data sample densities 
are roughly $\sim 0.01 \, {\rm GeV}$, $\sim 0.002 \, {\rm GeV}$ and $ 0.1 \, {\rm GeV}$ in the near resonances, 
resonances located and away resonances regions, respectively. 
We take the data by interpolating with evenly distribution under the interval $0.01 \, {\rm GeV}$. 
With the above discussions, Eq. (\ref{eq:Fpi_DR2_tl_1}) is converted to 
\beq 
|\mathcal{F}_{\pi}(s)|^2 &=& \Theta(s_{\rm max} - s) \, \vert \mathcal{F}^{\rm data}_{\pi, {\rm Inter.}}(s) \vert^2 \nonumber\\
&+& \Theta(s - s_{\rm max}) \,  \vert \mathcal{F}_{\pi}^{\rm pQCD}(s) \vert^2 \,. 
\label{eq:Fpi_DR2_tl_2}
\eeq

\section{Perturbative QCD formulism}\label{sec:pQCD-form}

The accuracy of pQCD prediction of spacelike pion EM form factor is now up to twist four of both two-particle (2p) and three-particle (3p)
pion LCDAs and to NLO QCD radiation corrections of 2p-to-2p scattering \cite{Cheng:2019ruz}, 
\beq 
{\cal F}_\pi^{\rm pQCD}(Q^2) &=& {\cal F}^{\rm pQCD}_{\pi, \rm{ t2}}(Q^2) + \mathcal{F}^{\rm pQCD}_{\pi, \rm{t3}}(Q^2) \nonumber\\
&+& {\cal F}^{\rm pQCD}_{\pi, \rm{t2 \otimes t4}}(Q^2) + {\cal F}^{\rm pQCD}_{\pi, {\rm 3p}}(Q^2) \,.
\label{eq:ff_pi}
\eeq 
The relation $Q^2 \equiv -q^2$ is implied for the spacelike form factor, 
and the contributions arose from leading twist, twist three and twist four LCDAs associated to 2p configuration, 
as well as the contribution from 3p configuration of LCDAs are shown separately in Eq. (\ref{eq:ff_pi}). 
The explicit expressions of the form factor accompanied with the 2p-to-2p and 3p-to-3p scattering are given in the supplemented materials. 
The dominate contributed terms associated with leading and subleading 2p LCDAs 
can be decomposed in terms of the lowest two Gegenbauer moments ($a_0^\pi = 1$ and $a_2^\pi$) and the chiral mass as 
\beq
&~&{\cal F}^{\rm pQCD}_{\pi, \rm{t2}}(Q^2) = \mathcal{F}_1^{\rm t2}(Q^2) + a_2^\pi \, \mathcal{F}_2^{\rm t2}(Q^2) + (a_2^\pi)^2 \, \mathcal{F}_3^{\rm t2}(Q^2) \,,
\label{eq:ff_t2_a2m0pi} \\
&~&{\cal F}^{\rm pQCD}_{\pi, \rm{t3}}(Q^2)= (m_0^{\pi})^2 \mathcal{F}_1^{\rm 2p, t3}(Q^2) + m_0^{\pi}  \mathcal{F}_2^{\rm 2p, t3}(Q^2) \nonumber\\ 
&~& \hspace{1.62cm} + (a_2^\pi)^2  \mathcal{F}_3^{\rm 2p, t3}(Q^2) +  a_2^\pi \mathcal{F}_4^{\rm 2p, t3}(Q^2) \nonumber\\ 
&~& \hspace{1.62cm} + m_0^{\pi} a_2^\pi  \mathcal{F}_5^{\rm 2p, t3}(Q^2) + \mathcal{F}_6^{\rm 2p, t3}(Q^2) \,,
\label{eq:ff_t3_a2m0pi} \\
&~&{\cal F}^{\rm pQCD}_{\pi, \rm{t2 \otimes t4}}(Q^2) = \mathcal{F}_1^{\rm 2p, t2 \otimes t4}(Q^2) + a_2^\pi \, \mathcal{F}_2^{\rm 2p, t2 \otimes t4}(Q^2) \,. 
\label{eq:ff_t2t4_a2} 
\eeq
Timelike form factor in the pQCD prediction has the similar accuracy and decomposition as in the spacelike one. 

\begin{figure}[b]
\vspace{-2mm}
\begin{center}
\includegraphics[width=0.48\textwidth]{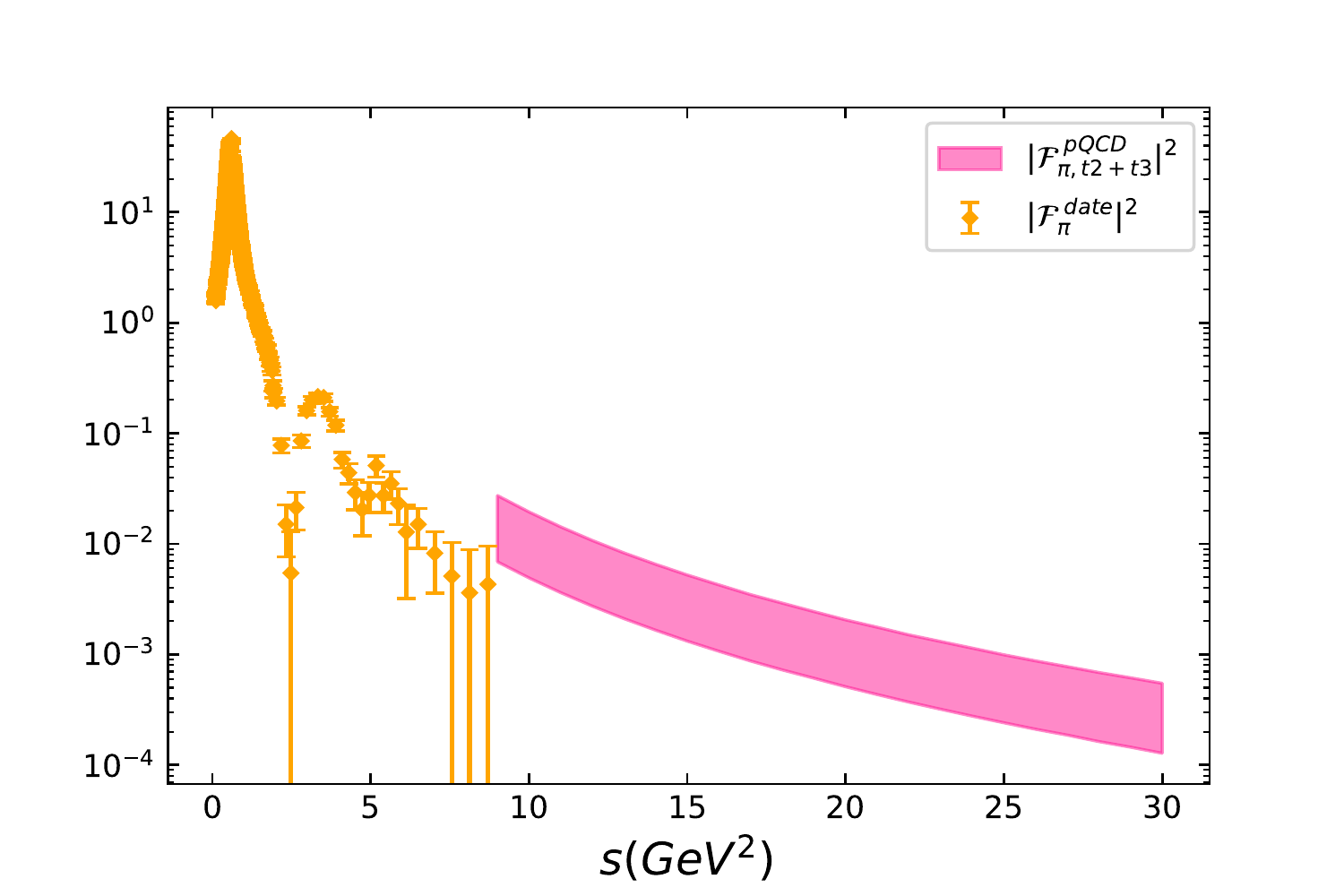}\non
\vspace{4mm}
\includegraphics[width=0.48\textwidth]{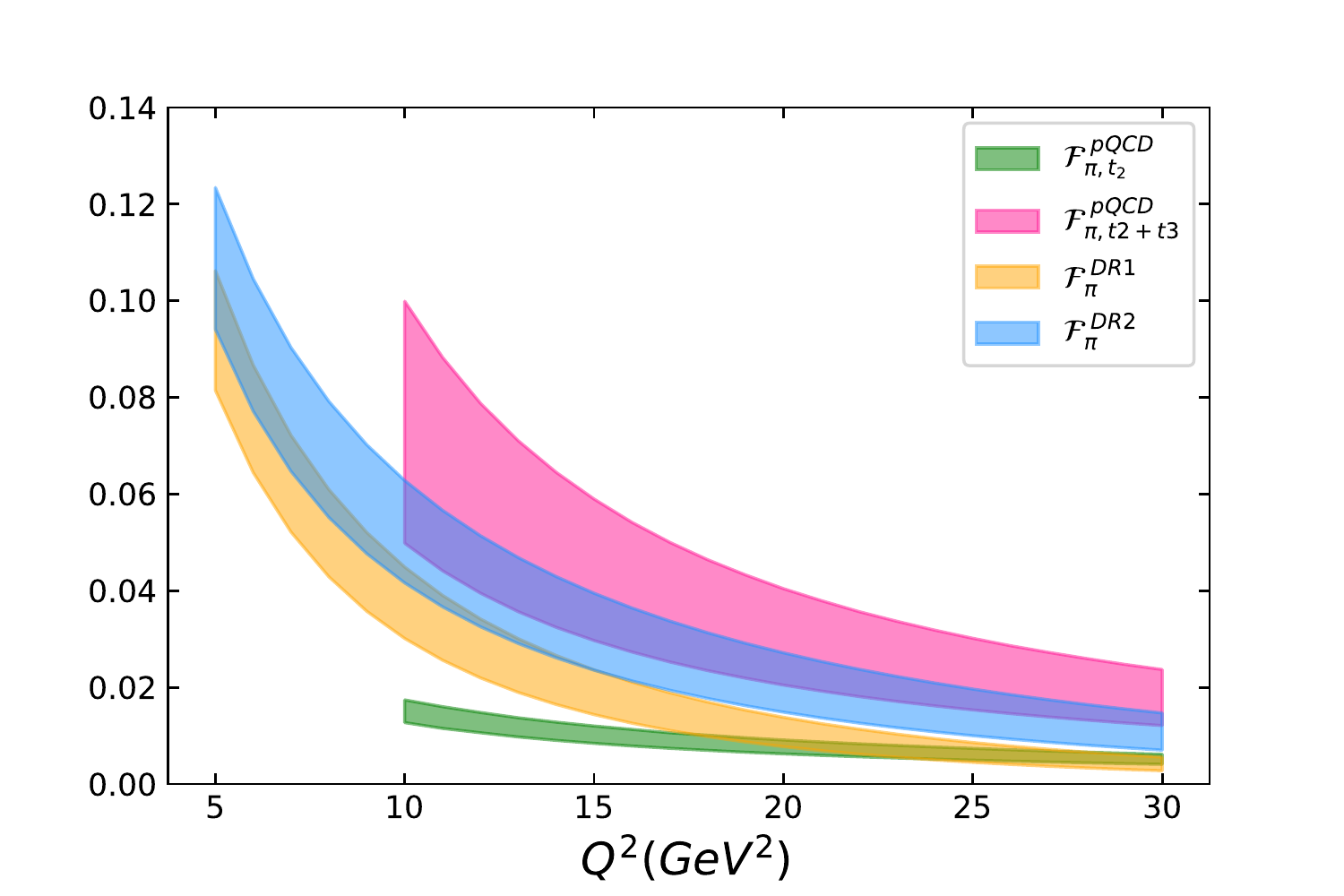}   
\end{center}
\vspace{-2mm}
\caption{Up panel: Timelike form factor with the data samples \cite{BaBar:2012bdw} in $s \in [4m_\pi^2, s_{\rm max} \simeq 8.7] \, {\rm GeV}^2$ 
and the PQCD prediction up to ${\rm 2p}$ twist three level in $[9, 30] \, {\rm GeV}^2$.  
Low panel: Spacelike form factor obtained from the modified dispersion relation 
and from the direct pQCD calculations in $Q^2 \in [10, 30] \, {\rm GeV}^2$. 
The testing pQCD predictions in both plots are obtained by taking the priori choice of $m_0^\pi$ and $a_2^\pi$.}
\label{fig1}
\end{figure}  

In the up panel of figure \ref{fig1}, we depict the timelike form factor measured in the $e^+e^-$ annihilation by BABAR collaboration, 
and also the high energy tail (magenta band) from the dominate pQCD prediction 
$\vert {\cal F}_{\pi, {\rm t2+t3}}^{\rm pQCD}(s) \vert = \vert {\cal F}^{\rm pQCD}_{\pi, \rm{t2}}(s) + {\cal F}^{\rm pQCD}_{\pi, \rm{t3}}(s) \vert$. 
In the pQCD evaluation, we have taken the priori choice of the parameters $m_0^\pi(1 \, {\rm GeV}) = 1.6 \pm 0.4 \, {\rm GeV}$ 
and $a_2^\pi(1 \, {\rm GeV}) = 0.25 \pm 0.25$, and take in to account the effect from the scale evolutions of these parameters. 
It is shown that the pQCD calculation marries with the BABAR data at the intermediate regions with in the uncertainty and error regions. 
In the low panel of figure \ref{fig1},  
we depict the spacelike form factor obtained by the modified dispersion relation in Eq. (\ref{eq:Fpi_DR2}). 
The full result ${\cal F}_\pi^{\rm DR2}(Q^2)$ with considering the high energy tail from pQCD evaluation is shown by the blue band, 
and the partial result ${\cal F}_\pi^{\rm DR1}(Q^2)$ without considering the high energy tail is shown by the orange one. 
It is found that the high energy tail gives a nonnegligible contribution, especially in the large momentum transfer regions. 
This could be traced to the logarithm expression of the timelike form factor in Eq. (\ref{eq:Fpi_DR2}) 
which strengthens the role of high energy tail in the dispersion relation. 
For the sake of comparison, we also depict the direct pQCD calculations at the leading power (cyan band) and 
up to ${\rm 2p}$ twist three level (magenta curve) with the priori choice of the parameters, 
the errors also come from the uncertainties of parameters $m_0^\pi$ and $a_2^\pi$ and mainly from the $m_0^\pi$. 
This testing plot from one side reveals that the chiral enhancement effect from ${\rm 2p}$ twist three LCDA is apparent in pion form factor. 
From another side, it shows that the uncertainty of the partial result obtained with only the BABAR data 
is larger than the leading twist contribution ${\cal F}^{\rm pQCD}_{\pi, \rm{ t2}}$ predicted by pQCD. 
In light of this point, the fit of Eq. (\ref{eq:Fpi_DR2}) with the direct pQCD calculation of spacelike form factor 
can not arrive at a good result for $a_2^\pi$ with well controlled errors. 

\section{Result and discussion}\label{sec:fit}

In this section we do the fit of spacelike form factor obtained from the modified dispersion relation and from the direct pQCD calculation. 
The minimal $\chi^2$ fitting is done in the large momentum transfer regions $10 \leq Q^2 \leq 30 \, {\rm GeV}^2$ 
where the pQCD calculation is applicable. 
\beq
\chi^2 = \sum_{i=1}^{11} \frac{\left[ {\cal F}_\pi^{\rm DR2}(Q_i^2) - {\cal F}_\pi^{\rm pQCD}(Q_i^2) \right]^2}{\left[ 
\delta {\cal F}_\pi^{\rm DR2} (Q_i^2) \right]^2} \,.
\label{eq:fit-chi2}
\eeq
The are two sources for the uncertainty of the dispersion relation deduced result, saying $\delta {\cal F}_\pi^{\rm DR2}$ in the denominator, 
one is the experimental data errors and the other one is the pQCD evaluation of high energy tail in the integrand of dispersion relation. 
We still use the broad intervals of $m_0^\pi$ and $a_2^\pi$ as in the previous test to estimate the later part. 
We take eleven energy points staring from $Q^2 = 10 \, {\rm GeV}^2$ with the step width $2 \, {\rm GeV}^2$. 
The full pQCD prediction in Eq. (\ref{eq:ff_pi}) is rearranged in terms of the parameters $m_0^\pi$ and $a_2^\pi$ 
at default scale $\mu_0 = 1 \, {\rm GeV}$ as 
\beq
{\cal F}_\pi^{\rm pQCD}(Q^2) &=& (m_0^\pi)^2 F_1(Q^2) + m_0^\pi F_2(Q^2) + F_3(Q^2) \\
&+& m_0^\pi a_2^\pi F_4(Q^2) + a_2^\pi F_5(Q^2) + (a_2^\pi)^2 F_6(Q^2) \,, \nonumber
\label{eq:pqcd-paras}
\eeq
in which the function $F_3(Q^2)$ collects the contributions from the asymptotic term and partial high twists terms in Eq. (\ref{eq:ff_pi}). 

\begin{table*}[t]
\begin{center}
\caption{Fitting results of $m_0^\pi$ and $a_2^\pi$ at the default scale $1 \, {\rm GeV}$. 
Scenario I (II) represents the fit with(out) considering the scale running of nonperturbative parameters in pQCD calculation, 
scenario IIA (IIB) indicates the fit with varying down (up) the renormalization and factorization scales by $25\%$ 
in contrast to the conventional one taken in scenario II.}  
\vspace{2mm}
\begin{tabular}{l | c c | c c }
\toprule
{\rm Scenario} \qquad\quad & \qquad\quad ${\rm I} \qquad\quad $ & \qquad\quad  ${\rm II}$ \qquad\quad & \qquad\quad ${\rm IIA}$ \qquad\quad & \qquad\quad  ${\rm IIB}$ \qquad\quad   \\
\hline
$m_0^\pi({\rm GeV})$    & $1.37^{+0.29}_{-0.32}$   & $1.31^{+0.27}_{-0.30}$ & $0.93^{+0.24}_{-0.27}$   & $1.59^{+0.30}_{-0.34}$ \\ 
$a_2^\pi$                       & $0.25 \pm 0.25$              & $0.23 \pm 0.25$            & $0.25 \pm 0.25$             & $0.26 \pm 0.25$ \\ 
\toprule
\end{tabular}
\vspace{-2mm}
\label{tab1}
\end{center}
\end{table*}

For the second moment of pion distribution amplitude, 
the first lattice calculation by using the momentum smearing technique shows $a_2^\pi(1 \, {\rm GeV}) = 0.135 \pm 0.032$ 
with full control of all systematic errors \cite{RQCD:2019osh}, 
while the very recent lattice result by using large-momentum effective theory shows a noticeably larger value $0.258^{+0.070}_{-0.052}$ \cite{Hua:2022kcm}. 
It is also studied by the global PQCD fit at leading order (LO) with considering the well-explained hadronic two-body $B$ decays \cite{Hua:2020usv}, 
nevertheless, the result shows obvious difference with that obtained from QCD sum rules (QCDSRs) \cite{Ball:2006wn}, dispersion derivation \cite{Li:2022qul} and LQCD evaluations. 
We mark that the result obtained from the fit of $B$ decays would firstly suffer large uncertainty from the inverse moment of $B$ meson $\lambda_B$, 
and secondly the fit is carried out at LO without considering the NLO corrections and power suppressed contributions. 
As stated in the last section, the dispersion relation deduced result with visible uncertainty could not figure out $a_2^\pi$ with well controlled uncertainty, 
so hereafter we take the priori result $a_2^\pi(1 \, {\rm GeV}) = 0.25 \pm 0.25$ as a constraint to fit it and $m_0^\pi$. 

\begin{figure}[b]
\vspace{-2mm}
\begin{center}
\includegraphics[width=0.48\textwidth]{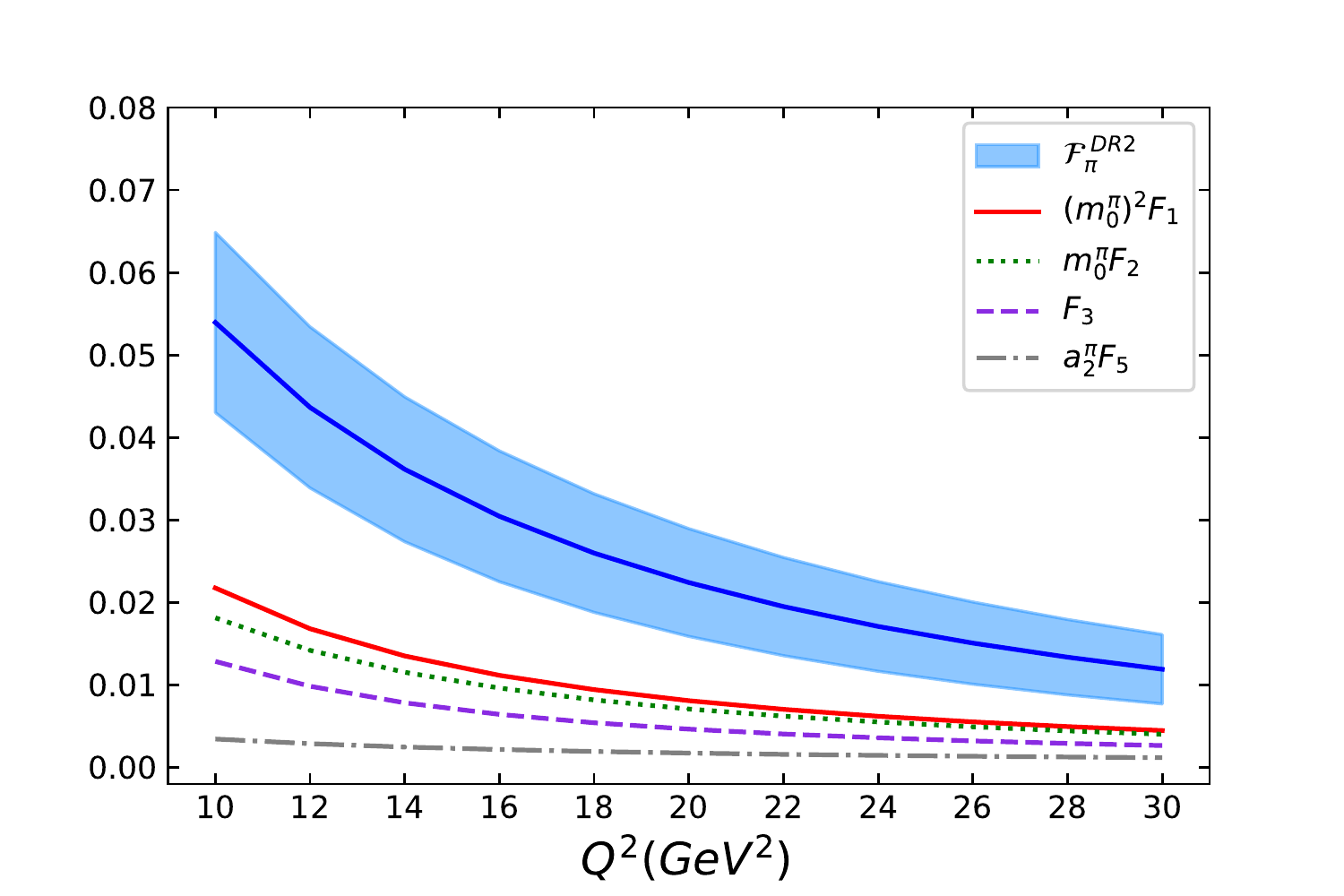} \non
\vspace{4mm} 
\includegraphics[width=0.48\textwidth]{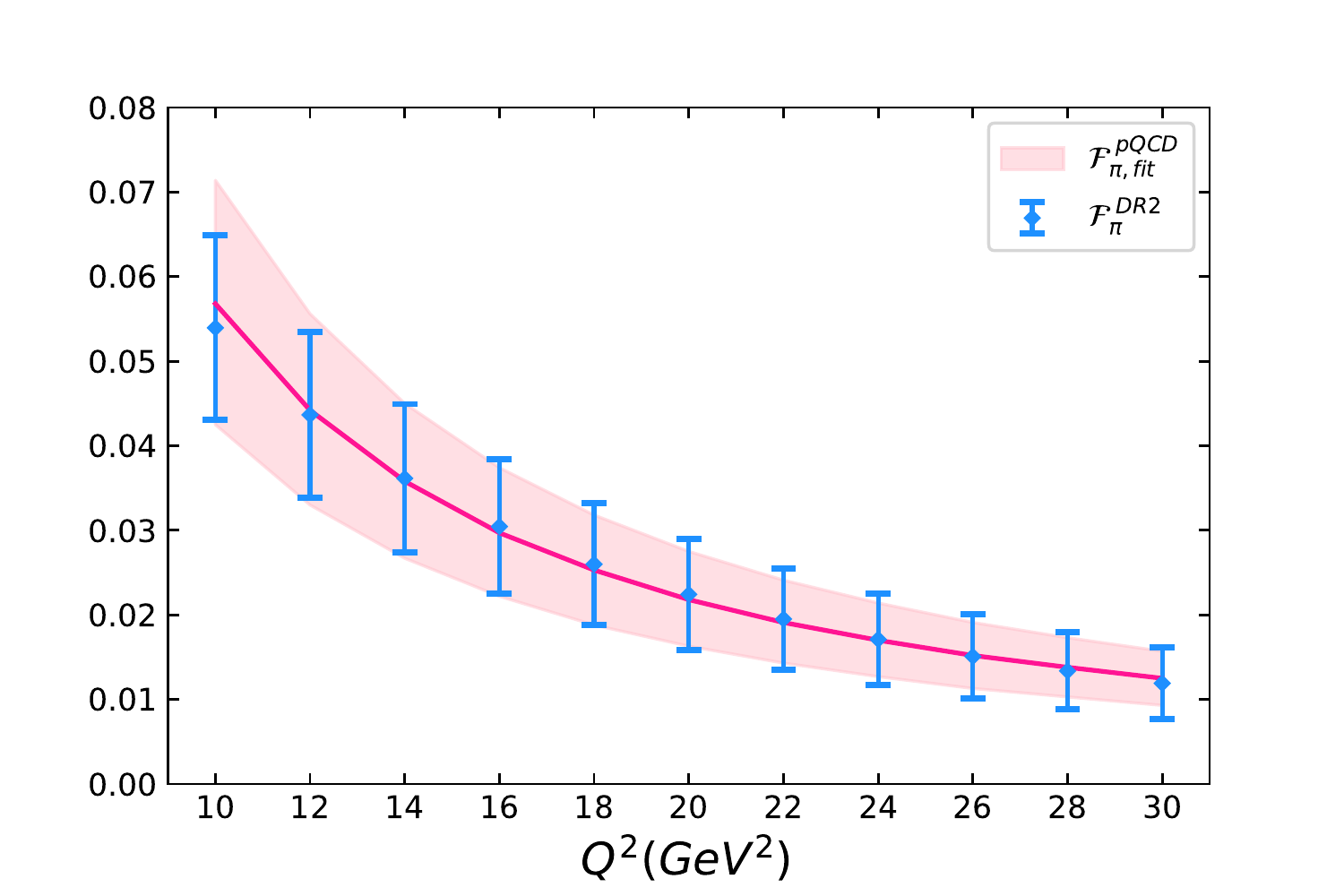}
\end{center}
\vspace{-4mm}
\caption{The spacelike form factor obtained from the modified dispersion relation and the pQCD calculation with the new obtained parameters in scenario II. 
In the upper panel, different terms in pQCD calculation are shown for the sake of comparison.}
\label{fig2}
\end{figure}  

The fit results of parameters $m_0^\pi$ and $a_2^\pi$ at default scale $1 \, {\rm GeV}$ are listed in table \ref{tab1}.  
We do the fit in two scenarios, the scenario ${\rm I}$ acquiesces the fixed scale $1 \, {\rm GeV}$ for the nonperturbative parameters in pion LCDAs, 
and in scenario ${\rm II}$ we consider the scale evolution of parameters in the pQCD evaluations. 
The factorization scale in pQCD calculation of hadron matrix element is formerly chosen at the largest internal virtuality. 
In order to examine the influence from the choice of factorization and renormalization scales, we vary it down and up by $25 \%$ 
respecting to the conventional one in the scenario ${\rm II}$. 
The notations ${\rm IIA, IIB}$ indicate the fit result obtained by taking the scales in pQCD calculation at 
$3\mu_t/4$ and $5\mu_t/4$, here $\mu_t$ is the hard scale in the scattering and also the scales taken in scenario ${\rm II}$. 
We see that (a) the scale running of the nonperturbative parameters does not bring significant modification to the fit result, 
indicating that fixing them at the default scale in previous pQCD calculations is a reasonable treatment, 
(b) the variation of scale choice brings about $20 \%$-$30 \%$ modification to the result of $m_0^\pi$, 
which reveals the possible nonnegligible correction from the next-to-next-leading-order QCD correction to the form factor, 
(c) the fit result of $m_0^\pi$ is well under control with the current data accuracy, 
and the data-driven approach developed here to extract the nonperturbative parameters does not rely too much on the choice of factorization scale. 

\begin{widetext}
\begin{figure*}[t]
\vspace{-2mm}
\begin{center}
\includegraphics[width=0.8\textwidth]{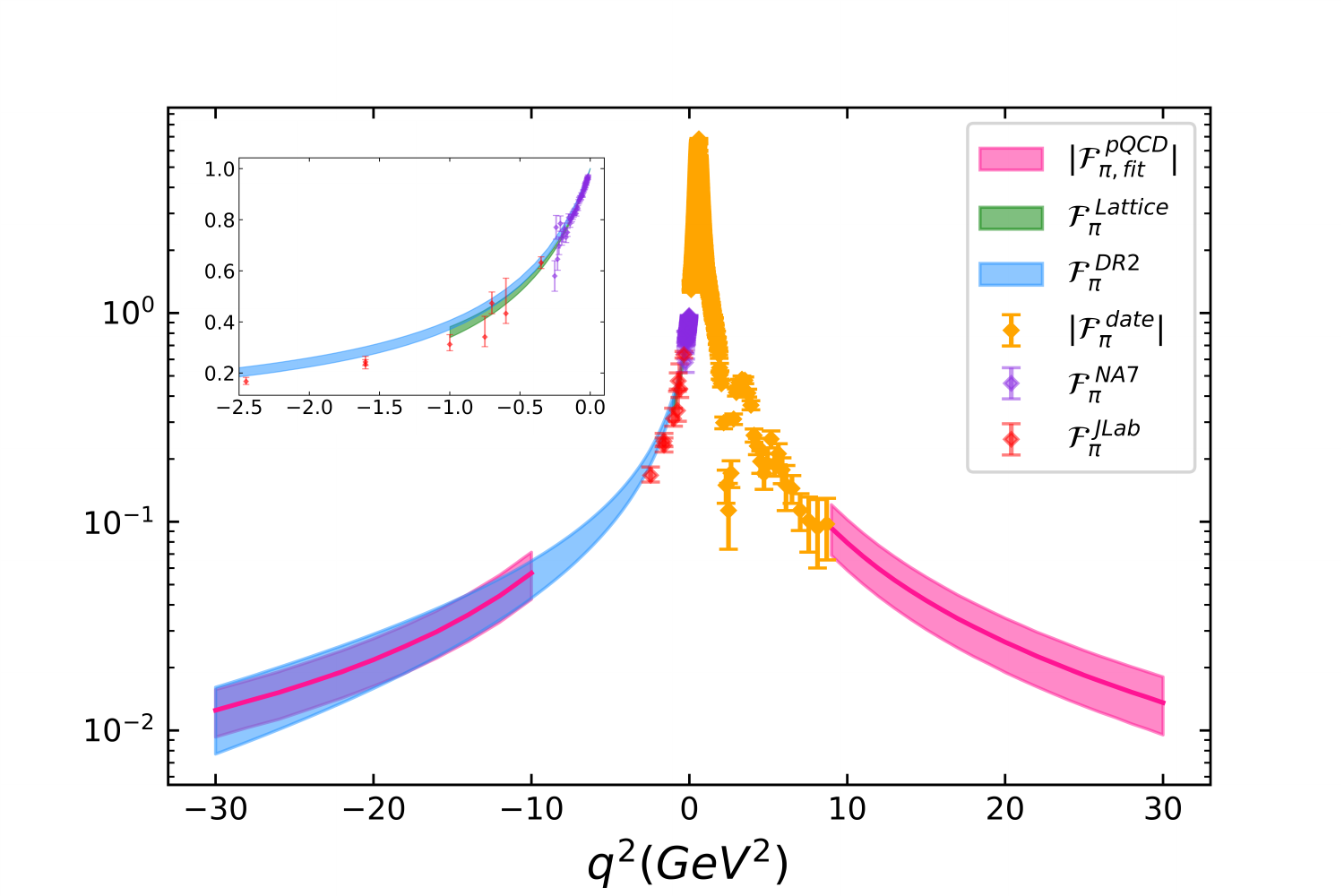}  
\end{center}
\vspace{-4mm}
\caption{The magnitude of pion electromagnetic form factor in the whole kinematic regions.}
\label{fig3}
\end{figure*} 
\end{widetext}

We depict in figure \ref{fig2} the spacelike form factor obtained from the modified dispersion relation 
and from the pQCD calculation ${\cal F}^{\rm pQCD}_{\pi, {\rm fit}}$ with the new fit parameters in the scenario II. The plot in the up panel presents the shape curves of pQCD functions associated with the parameters in our interesting, as shown in Eq. (\ref{eq:pqcd-paras}).
The curves of functions $m_0^\pi a_2^\pi F_{4}(Q^2)$ and $(a_2^\pi)^2 F_{6}(Q^2)$ are not presented since they are very close to zero. 
The plot in the low panel shows the pQCD result with the new fitted parameters, in contrast to the dispersion relation deduced result.
In figure \ref{fig3} we plot the form factor in the whole momentum transfer/invariant mass regions, 
where the direct measurements available in the regions $q^2  \in [4m_\pi^2, \simeq 8.70] \, {\rm GeV}^2$ \cite{BaBar:2012bdw}, 
the dispersion relation deduced result and the pQCD prediction in the region with large $\vert q^2 \vert$ 
are depicted by orange data-bar, blue and magenta bands, respectively. 
In the later two result, the new obtained parameters in the scenario II are employed. 
We can see that the new pQCD prediction consists with BABAR data in the intermediate region much better than the testing pQCD calculation as shown in figure \ref{fig1}.
Moreover, we present the direct measurement of spacelike form factor from ${\rm NA7}$ \cite{NA7:1986vav} and Jefferson Lab $F_\pi$ collaborations  \cite{JeffersonLabFpi-2:2006ysh,JeffersonLab:2008jve} in the region $q^2 \in [-2.5, 0] \, {\rm GeV}^2$, 
and also the precise LQCD evaluation (Green band) carried out in the large recoil regions $[-1, 0] \, {\rm GeV}^2$ \cite{Wang:2020nbf},
this piece is enlarged and embodied in the top-left corner of the figure. 
It is shown a good agreement in the large recoiled region $q^2 \in [-1, 0] \, {\rm GeV}^2$ within uncertainties and experimental errors. 
Whereas in the region $-2.5 \, {\rm GeV}^2 \leq q^2 \lesssim -1.0 \, {\rm GeV}^2$, 
the dispersion relation deduced form factor lies slightly above the experimental points, 
this inconsistency is also happened in the LCSRs study where the high energy tail is parameterized by the duality resonant model \cite{Cheng:2020vwr}. 
The Jefferson Lab $12 \, {\rm GeV}$ program would say more in the intermediate momentum transfer regions. 

\section{Summary}\label{sec:summary}

Inspired by the precise measurement of pion EM form factor in the resonant regions which marks up the missing piece without QCD-based calculation, 
we study the form factor with the modulus squared dispersion relation in which the line shapes in the large invariant mass/momentum transfer regions 
are calculated from pQCD approach. 
The main target of this work is to extract the most important nonperturbative parameters in pion LCDAs, 
saying the chiral mass and the second Gegenbauer moment, 
which is usually taken from other approaches as inputs in the pQCD calculation and result in a large uncertainty for the prediction. 
With fitting the spacelike form factor obtained from the modulus squared dispersion relation and from the direct pQCD calculation, 
we obtain the result $m_0^\pi = \left(1.31^{+0.27}_{-0.30} \right) \, {\rm GeV}$ and $a_2^\pi = 0.23 \pm 0.26$ at the default scale $1 \, {\rm GeV}$, 
this result is consist with previous choice in pQCD calculation. 
In order to examine the influence from the choice of factorization and normalization scales, 
we vary them by $25 \%$ respecting to the conventional one and find that 
the data-driven method formulated in this work is robust to extract the nonperturbative parameters. 
There are two directions we can strive for in this research. 
The first one is to increase the accuracy of measurement, especially in the region closing to $s_{\rm max}$, 
to reduce the uncertainty of spacelike form factor obtained from the dispersion relation, 
and hence improve the ability of this method to extract the second Gegenbauer moment $a_2^\pi$. 
Secondly, apply this method into other processes involved pion, like the pion transition form factor, 
to find the optimal choice of factorization and normalization scales in pQCD approach with the combined analysis. 

\section{Acknowledgements}\label{sec:summary}

We would like to thank Guang-shun Huang and Gen Wang for the useful discussions on the experiment measurement and lattice evaluation, respectively, 
especially to Hsiang-nan Li for the careful reading of the draft and the fruitful comments.
This work is supported by the National Science Foundation of China (NSFC) under the Grants No. 11975112 
and the Joint Large Scale Scientific Facility Funds of the NSFC and CAS under Contract No. U1932110 and U2032102. 
S.C. is also supported by the Natural Science Foundation of Hunan Province, China (Grant No. 2020JJ4160).


\clearpage

\newpage


\begin{widetext}

\begin{center}
{\large Supplement materials: Pion electromagnetic form factors from the perturbative QCD approach}
\end{center}
 
\begin{figure}[thb]
\vspace{4mm}
\begin{center}
\includegraphics[width=0.8\textwidth]{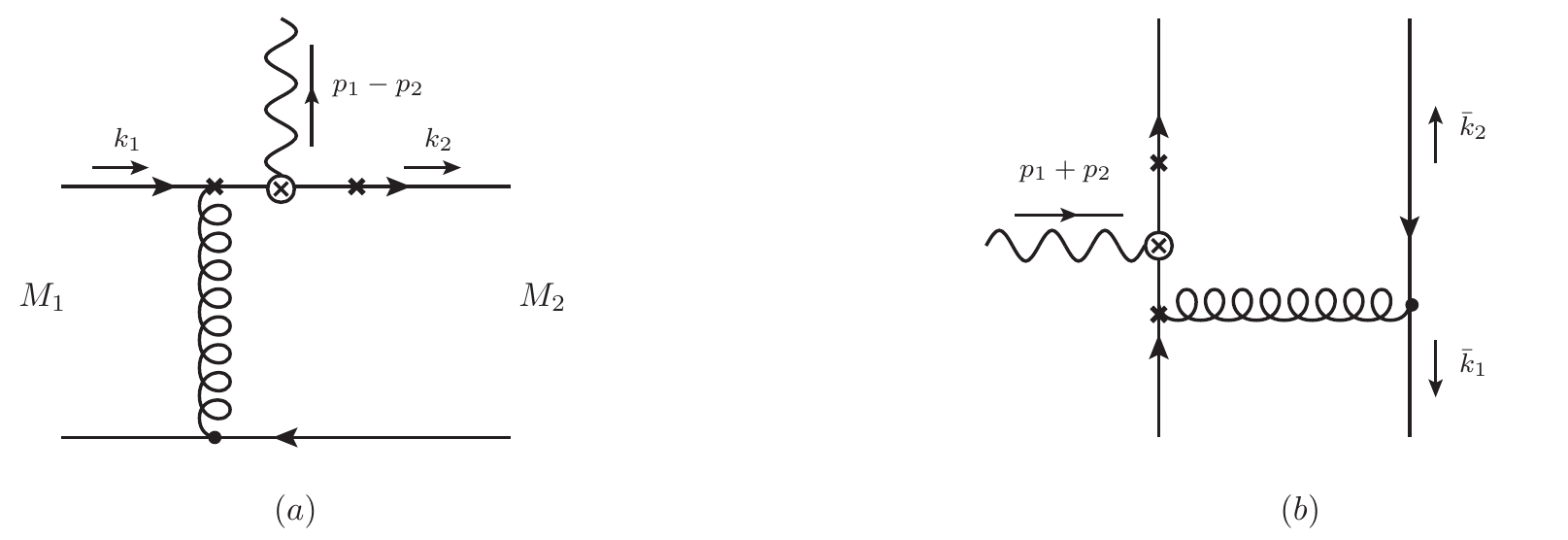}
\end{center}
\vspace{-4mm}
\caption{Feynman diagrams of spacelike (left) and timelike (right) pion form factors at leading-order.}
\label{fig4}    
\end{figure}     

We present the Feynman diagrams of pion EM form factors at leading order (LO) in Fig. \ref{fig4}. 
The EM vertexes denoted by $\otimes$ are separately shown in the down and up quark lines, 
and the sum of these two diagrams gives the invariant amplitudes of transitions with 
EM current $j_\mu^{{\rm em}} = \left( {\bar u} \gamma_\mu u - {\bar d} \gamma_\mu d \right)$. 
The possible attachments of internal hard gluons are denoted by $\times$. 
In our kinematics, $p_1 = (Q/\sqrt{2}, 0, {\bf 0}), p_2 = (0,Q/\sqrt{2}, {\bf 0})$ are momenta of external mesons, 
$k_1 = (xQ/\sqrt{2}, 0, {\bf k}_{\rm 1T})$ and $k_2 = (0,yQ/\sqrt{2}, {\bf k}_{\rm 2T})$ are the momentum carried by the quark lines in the external mesons, 
with $x$ and $y$ (${\bf k}_{\rm 1T}$ and ${\bf k}_{\rm 2T}$), respectively, being the longitudinal momentum fractions (transversal momentum), 
the momentum transfer squared is $q^2 = (p_1-p_2)^2$ in the spacelike case, while it is $q^2 = (p_1+p_2)^2$ in the timelike case.

We take the spacelike pion form factor for example to explain the basic idea of pQCD approach. 
It is defined by the nonlocal matrix element 
\beq
\langle \pi^-(p_2) \vert j_{\mu, q}^{\rm em} \vert \pi^-(p_1) \rangle \equiv e_q (p_1 + p_2) {\cal F}_\pi(Q^2) \,,  
\label{eq:ff-1}
\eeq
here $j_{\mu, q}^{\rm em}$ is the $q$-flavor component in the EM current. 
With separating the contributions from short and long-distance interactions, the matrix element is written in the factorizable formulism as 
\beq
&&\langle \pi^-(p_2) \vert J_\mu^{\mathrm{e.m.}} \vert \pi^-(p_1) \rangle = \oint dz_1 dz_2 \, \big\langle \pi^-(p_2) \bigg\vert \left\{ \overline{d}_\gamma(z_2) \,
\mathrm{exp} \left( ig_s \int_{0}^{z_2} d\sigma_{\nu^\prime} A_{\nu^\prime}(\sigma) \right) u_\beta(0) \right\}_{kj} \bigg\vert 0 \big\rangle_{\mu_t}  \non
&&\hspace{3cm} H_{\gamma\beta\alpha\delta}^{ijkl}(z_2,z_1) \, \big\langle 0 \bigg\vert \left\{ \overline{u}_\alpha(0) \,
\mathrm{exp} \left(i g_s \int_{z_1}^{0} d\sigma_\nu A_\nu(\sigma) \right) d_\delta(z_1) \right\}_{il} \bigg\vert \pi^-(p_1) \big\rangle_{\mu_t} \, ,
\label{eq:ff_sl-fact}
\eeq
where $\gamma , \beta, \alpha, \delta$ are the spinor indices, $i,j,k,l$ are the color indicators, and $\mu_t$ is the factorizable scale. 
The large momentum transfer square $\vert q^2 \vert$ ensures the smallness of relative interaction distance $\overline{z}_1 - \overline{z}_2$, 
in this case $ \left( \overline{z}_1 - \overline{z}_2 \right) \cdot \overline{q} \sim 1$ and hence the expansion parameter for a given operator is the twist. 
In the factorisation fomulism Eq. (\ref{eq:ff_sl-fact}), the matrix elements on the RHS indicate the amplitudes of mesons breaking-up into partons, 
or inversely of the hadronization. It is can be rewritten in terms of contributions from different Gamma matrixes via the Fierz identity 
\beq
&&\big\langle 0 \bigg\vert \left\{ \overline{u}_\alpha(0) \,
\mathrm{exp} \left(i g_s \int_{z_1}^{0} d\sigma_\nu A_\nu(\sigma) \right) d_\delta(z_1) \right\}_{il} \bigg\vert \pi^-(p_1) \big\rangle_{\mu_t} \non
=&& \frac{\delta_{il}}{3} \left\{
\frac{1}{4} \left(\gamma_5 \gamma^\rho \right)_{\delta\alpha} \big\langle 0 \big\vert \overline{u}(0) \, \mathrm{exp} \left( i g_s \int_{z_1}^{0} \, 
d\sigma_\nu A_\nu(\sigma) \right) \left(\gamma_\rho \gamma_5 \right) d(z_1) \big\vert \pi^-(p_1) \big\rangle_{\mu_t} + \cdots \right\} \,.
\label{eq:Fierz}
\eeq
And the remaining nonlocal matrix elements with certain Gamma matrixes are usually expanded by the LCDAs at different twists, 
not only for the assignment of a pair of soft quarks (two particle configuration), 
but also for the three soft partons with an additional soft gluon (three particle configuration)\cite{Ball:2006wn}. 
The hard kernel $H$ in Eq.~(\ref{eq:ff_sl-fact}) is pQCD calculable. 
We here only show explicitly the general expression accompanied with two particle configuration  
\beq
H_{\gamma\beta\alpha\delta}^{ijkl}(z_1,z_2) = (-1) \left[ i g_s \gamma_m \right]_{\alpha\beta} T^{ij}
\left[(i e_q \gamma_\mu) S_0(0-z_1) (i g_s \gamma_n) \right]_{\gamma\delta} T^{kl}  \left[-i D_{mn}^0(z_1-z_2) \right] \,,
\label{eq:hk-2p}
\eeq
where the free propagators read as
\beq
S_0(z) = \frac{i}{2\pi}\frac{\zsl}{z^4} \,, \,\,\,\,\,\, D_{mn}^0(z) = \frac{1}{4\pi}\frac{g_{mn}}{z^2} \,.
\eeq

The pQCD result is quoted here separately for 2p-to-2p and 3p-to-3p scatterings as \cite{Cheng:2019ruz}
\beq
&~&{\cal F}^{\rm 2p}_\pi(Q^2) = \frac{8}{9} \alpha_s \pi f_\pi^2 Q^2 \int_0^1 dx \int_0^1 dy \, \int_0^{\frac{1}{\Lambda}} b_1 db_1 b'_1 db'_1 \, 
e^{-S_{\rm 2p}(x_i,y_i,b,b',\mu)} \Big\{ \bar{y} \varphi_\pi(x) \varphi_\pi(y) \, \Big[1+ F_{t2}^{(1)} (x, y, t, Q^2)\Big] \, \mathcal{H} \non
&~& + \frac{2m_0^2}{Q^2} \Big[- y \varphi_\pi^p(x) \varphi_\pi^p(y) \left[1+F_{t3}^{(1)} (x, y, t, Q^2)\right]  \, \mathcal{H} 
+ \frac{1}{6} \varphi_\pi^p(x) \varphi_\pi^\sigma(y) \left[-y Q^2\, \mathcal{H}_1
-(\bar{x}-\bar{y}+2\bar{y}-x\bar{y})Q^2 \, \mathcal{H}_2  - \mathcal{H}\right]  \Big] \, S_t(\bar{y})\non
&~& + 2 \Big[ g_{2\pi}(x) \varphi_\pi(y) \, \bar{x}\bar{y}  \mathcal{H}_2
+ \varphi_\pi(x) g_{2\pi}(y)  \, \bar{y}^2 \left[\mathcal{H}_1 + \mathcal{H}_2 \right] \non
&~& \hspace{0.6cm} + \big[ \varphi_\pi(x) g_{1\pi}(y)  - \varphi_\pi(x) g_{2\pi}^{\dag}(y) \big]
\left[ 2 \bar{y} (\mathcal{H}_1 + \mathcal{H}_2 + \bar{y}(2-x) \mathcal{H}_3) \right]\Big]  \, S_t(\bar{y})\Big\} \,,
\label{eq:ff-2p} \\  
&~&{\cal F}^{\rm 3p}_\pi(Q^2) = \frac{16}{3} \alpha_s \pi f_{\pi}^2 Q^2 \int_0^1 \mathcal{D}x_i \int_0^1 \mathcal{D}y_i \,
\int_0^{\frac{1}{\Lambda}} \, b_1 db_1 b'_1 db'_1 b_2^2 db_2 \, e^{-S_{\rm 3p}(x_i,y_i,b_i,\mu)} 
\Big\{ \frac{f_{3\pi}^2}{f_{\pi}^2} (1-y_1) \varphi_{3\pi}(x_i) \varphi_{3\pi}(y_i) \, \mathcal{H}' \non
&~& + \frac{1}{2Q^2} \Big[ \varphi_\parallel^\dag(x_i) \varphi_\parallel^\dag(y_i)
\big[ \left(-4 (1-y_1)+ (1-y_1)y_2 \right) Q^2 \, \mathcal{H}'_2 + 5(1-y_1)y_2Q^2 \,\mathcal{H}'_3 \big] \non
&~&\hspace{0.7cm} + \varphi_\parallel^\dag(y_i) \varphi_\perp(x_i)  \big[ 4\,\mathcal{H}' +(1-y_1)Q^2 \, \mathcal{H}'_1 - (1-x_1)(1-y_1)Q^2 \, \mathcal{H}'_2 + (1-x_1)(1-y_1)x_2 Q^2 \, \mathcal{H}'_3 \big] \non
&~& \hspace{0.7cm} + \varphi_\perp(y_i) \varphi_\parallel^\dag(x_i)  \big[ -y_1(1-y_1)Q^2\, \mathcal{H}'_2 + y_1 (1-y_1) y_2Q^2 \, \mathcal{H}'_3 \big] 
+ \varphi_\perp(y_i) \varphi_\perp(x_i)   5y_1 \, \mathcal{H}' + [\varphi \rightarrow \tilde{\varphi} ]  \Big] \Big\} \,.
\label{eq:ff-3p}
\eeq
Here $\varphi_\pi$, $\varphi_\pi^{p/\sigma}$ and $g_{2\pi}$ denote the twist-2, twist-3 and twit-4 LCDAs of pion meson with two particle configuration, 
$\varphi_{3\pi}$ and $\varphi_{\parallel,\perp}, \tilde{\varphi}_{\parallel,\perp}$ are the twist-3 and twist-4 LCDAs associated to three particle configurations, respectively, $\varphi_{\parallel,\perp}^\dag$ are the auxiliary DAs related to $\varphi_{\parallel,\perp}$. 
What's more, $b_i$ is the conjugate transversal extent to ${\bf k}_{{\rm iT}}$, $S_{\rm 2p}$ and $S_{\rm 3p}$ are the $k_T$ Sudakov suppressed functions, 
$S_t$ is the threshold Sudakov function start to be appeared at subleading twist due to the skewed distribution of $\varphi_\pi^{p/\sigma}$, 
and the hard functions $\mathcal{H}$ indicate the Fourier integral from ${\bf k}_{{\rm iT}}$ to $b_i$ which is usually written in the product of Bessel functions. 
$F_{t2}^{(1)}$ and $F_{t3}^{(1)}$ are the NLO corrections functions associated to ${\rm 2p}$ twist-2 and twist-3 LCDAs \cite{Li:2010nn,Cheng:2014gba}, respectively.

We have truncated the gegenbauer expansion of leading twist LCDAs to the second order, 
and taken the ${\rm 2p}$ twist three LCDAs up to NLO in conformal spin and the second moments in truncated conformal expansion. 
\beq
\varphi_\pi(x, \mu)& = &6 x (1-x) \left[ 1 + a_2(\mu) C_2^{3/2}(2x - 1) 
 \right] \,,
\label{eq:DA-t2} \\
\varphi^{ p}_\pi(x, \mu) &=& 1 + 3 \rho_{\pi} \Big( 1 +6a_2 \Big)(1+\ln x)
- \frac{\rho_{\pi}}{2} \Big( 3 +54a_2 \Big) \, C_1^{1/2}(2x-1) \non
&+& 3 \Big(10 \eta_{3 \pi} +5 a_2  \rho_{\pi} \Big) \, C_2^{1/2}(2x-1)
-  \frac{9}{2} a_2  \rho_{\pi}   \, C_3^{1/2}(2x-1) - 3 \eta_{3\pi} \omega_{3\pi} \, C_4^{1/2}(2x-1)   \, , \label{eq:P-DA-t3-P}\\
\varphi_\mathcal{P}^\sigma(x, \mu) &=& 6x(1-x) \Big\{ 1 + \frac{\rho_{\pi}}{2} \Big(2 + 30 a_2(\mu)\Big)
-  \frac{15}{2}a_2(\mu) \rho_{\pi} \, C_1^{3/2}(2x-1) \non
&+& \frac{1}{2} \Big( \eta_{3 \pi}(10-\omega_{3 \pi}) + 3\rho_{\pi} a_2(\mu) \Big) \, C_2^{3/2}(2x-1)
+ 3 \rho_{\pi}\Big( 1 + 6 a_2(\mu) \Big) \, \ln x \Big\} \,.
\label{eq:DA-t3-T} 
\eeq
Two particle twist-3 DAs relate to both the leading twist DA $\varphi_\pi$ and the ${\rm 3p}$ twist three DAs, 
whose contributions are separated clearly in the above equations. 
The parameters $f_{3\pi}$ and $\omega_{3 \pi}$ can be defined by the matrix element of local twist-3 operators \cite{Ball:2006wn}, 
and $\eta_{3\pi}=f_{3\pi}/(f_\pi m_0^\pi)$,$\rho_{\pi}=m_{\pi}/m_0^{\pi}$.
We note that the scale dependences of parameters $a_2$ and $m_0^\pi$ \cite{Ball:2006wn} have been absorbed into 
the expansion functions ${\cal F}_{i=1-3}^{\rm t2}$ and ${\cal F}_{i=1-6}^{\rm 2p, t3}$, 
and the parameters in Eqs. (\ref{eq:ff_t2_a2m0pi},\ref{eq:ff_t3_a2m0pi},\ref{eq:ff_t2t4_a2},\ref{eq:pqcd-paras}) are conventionally set at $1 \, {\rm GeV}$. 
In our calculation we take $m_{\pi}=0.14 \, {\rm GeV}$, $f_{\pi}=0.13  \, {\rm GeV}$, $f_{3\pi}=0.0045\pm 0.0015 \, {\rm GeV}^2$ and $\omega_{3\pi}=-1.5\pm 0.7$. 
The Gegenbauer polynomials are  
\beq
C_1^{1/2}(t)= t, &\qquad& C_1^{3/2}(t)= 3t \non
C_2^{1/2}(t)=\frac{1}{2}(3t^2-1), &\qquad& C_2^{3/2}(t)=\frac{3}{2}(5t^2-1) \non
C_3^{1/2}(t)=\frac{1}{2}[5t^3-3t], &\qquad& C_4^{1/2}(t)=\frac{1}{8}(35t^4-30t^2+3) \,,
\eeq
in which $t =2x -1$.

In our calculation we do not take into account the $u, d$ quark masses in ${\cal F}^{\rm 2p, t2 \otimes t4}_\pi(q^2)$ and ${\cal F}^{\rm 3p}_\pi(q^2)$, 
that's why the quark mass correction terms and the $m_0^{\pi}$ terms are not appeared in these pieces. 
Two particle twist four LCDAs appeared in the third term in Eq. (\ref{eq:ff_pi}) read as  
\beq
g_{2\mathcal{P}}(x) = -\frac{1}{2}\int_0^x dx' \psi_{4\mathcal{P}}(x') \,, \quad\quad
g_{1\mathcal{P}}(x) = \frac{1}{16} \phi_{4\mathcal{P}}(x) + \int_0^x dx' g_{2\mathcal{P}}(x') \,, 
\label{eq:DA-t4-2p}
\eeq
in which the corrected expressions are \cite{Khodjamirian:2009ys}
\beq
&&\psi_{4\pi}(x) = \delta_\pi^2 \Big[ \frac{20}{3} \, C_2^{1/2}(2x-1) \Big] \,,
\label{eq:DA-t4-t2-psi} \\
&&\phi_{4\pi}(x) = \delta_\pi^2 \left[ \frac{200}{3}  x^2\bar{x}^2 
+ 21 \omega_{4\pi} \Big( x\bar{x} (2+13x\bar{x}) + [2x^3(6x^2-15x+10) \ln x] + [x \leftrightarrow \bar{x}] \Big) \right] \,. 
\label{eq:DA-t4-t2-phi}
\eeq
Here we only consider the contributions from the "genuine" ${\rm 3p}$ twist four DAs $\varphi_\parallel(x_i), \varphi_{\perp}(x_i)$, 
which are charactered by the parameters $\delta_\pi^2$, 
and the contributions arose from the Wandzura-Wilczek-type mass corrections are neglected since they are proportional to $m_\pi^2$. 

The conformal expansion of ${\rm 3p}$ twist three and twist four DAs read as
\beq
&&\varphi_{3\pi}(x_i) = 360 x_1 x_2 x_3^2 \left[1 + \frac{\omega_{3\pi} }{2} (7x_3-3) \right] \,,
\label{eq:DA-t3-3p}\\
&&\psi_\parallel(x_i) = 120x_1x_2x_3 \,\delta_{\pi}^2 \left[ \frac{21}{8}  \omega_{4\pi} (x_1-x_2) \right] \,, \quad 
\psi_\perp(x_i) = 30x^2_3 \, \delta_\pi^2 \left[ \frac{1}{3}(x_1-x_2) + \frac{21}{4} \omega_{4\pi}(x_1-x_2)(1-2x_3)\right] \,, 
\label{eq:DA-t2-3p-AV-perp} \\
&&\tilde{\psi}_\parallel(x_i) = -120x_1x_2x_3 \delta_\pi^2 \left[ \frac{1}{3} + \frac{21}{8}\omega_{4\pi}(1-3x_3) \right] \,, \quad
\tilde{\psi}_\perp(x_i) = 30x_3^2 \, \delta_\pi^2 \left[ \frac{1}{3}(1-x_3) +\frac{21}{4}\omega_{4\pi}(1-x_3)(1-2x_3) \right] \,,
\label{eq:DA-t2-3p-V-perp}
\eeq
in which one more parameters $\omega_{4 \pi}=0.20 \pm0.10$ is introduced. 
The auxiliary DAs $\varphi_\parallel^\dag(x_i)$ and $\varphi_\parallel^\dag(y_i)$ in Eq. (\ref{eq:ff-3p}) are defined by 
\beq
\varphi_\parallel^\dag(x_i) \equiv \int_0^{x_1} \, dx'_1 \, \varphi_\parallel(x'_1,x_2,x_3)\,,  \qquad
\varphi_\parallel^\dag(y_i) \equiv \int_0^{y_2} \, dy'_2 \, \varphi_\parallel(y_1,y'_2,y_3)\,
\eeq
with the bound conditions $\varphi_\parallel(x_1=0/1,x_2,x_3) = 0$ and $\varphi_\parallel(y_1,y_2=0/1,y_3)=0$.

 
\end{widetext}

\end{document}